\documentclass[fleqn,10pt]{wlscirep}
\usepackage{float}
\usepackage{caption}
\usepackage{subcaption}
\usepackage{upgreek}
\usepackage[utf8]{inputenc}
\usepackage[T1]{fontenc}

\newcommand{\pre}[1]{ \ensuremath{ \overrightarrow{#1} } }
\newcommand{\retro}[1]{ \ensuremath{ \overleftarrow{#1} } }

\newcommand{\expect}[1]{ \ensuremath{ \langle #1 \rangle } }
\newcommand{\qN}[1]{ \ensuremath{q^{(#1)}} }
\usepackage{graphicx}

\usepackage[normalem]{ulem}

\title{Measurement-based preparation of non-Markovian and multimode mechanical states}

\author[1]{Chao Meng}
\author[1,2]{George A. Brawley}
\author[1]{Soroush Khademi}
\author[1]{Elizabeth M. Bridge}
\author[1]{James S. Bennett}
\author[1,*]{Warwick P. Bowen}
\affil[1]{Australian Research Council Centre of Excellence for Engineered Quantum Systems, School of Mathematics and Physics,
University of Queensland, St Lucia, Queensland 4072, Australia}
\affil[2]{Terra 15, Level 9 / 256 Adelaide Terrace, Perth WA 6000, Australia}

\affil[*]{w.bowen@uq.edu.au}

\begin{abstract}

Nanomechanical resonators are a key tool for future quantum technologies such as quantum force sensors and interfaces, and for studies of macroscopic quantum physics. The ability to prepare room temperature non-classical states is a major outstanding challenge. Here, we explore the use of measurement-based state conditioning to achieve this. We demonstrate conditional cooling of a nanomechanical resonator that has non-Markovian decoherence, and show theoretically that the non-Markovianity makes quantum squeezing significantly easier to achieve. We further show that collective measurement of multiple resonator modes improves the quality of state preparation. This allows us to achieve collective thermomechanical squeezing, in experiments that go beyond the validity of the rotating-wave approximation. Our modelling shows that non-Markovianity and multimode conditioning can both enable room temperature quantum squeezing with existing technology.
Together, our results pave the way towards realising room temperature quantum nanomechanical devices and towards their application in quantum technology and fundamental science.

\end{abstract}
\begin{document}

\flushbottom
\maketitle
\thispagestyle{empty}

Quantum measurement is widely used to prepare non-classical states, and has important applications in quantum technologies from deterministic quantum computing~\cite{knill_scheme_2001,riste_deterministic_2013}, to quantum sensing ~\cite{cox_deterministic_2016,sayrin_real-time_2011} and fundamental tests of quantum mechanics~\cite{minev_catch_2019,brawley_nonlinear_2016}. A common scenario is continuous measurement of the position of a harmonic oscillator~\cite{aspelmeyer_cavity_2014,braginsky_quantum_1980}. In the usual operating regime, this localises the position and momentum equally~\cite{bowen_quantum_2015,doherty_quantum_2012,wilson_measurement-based_2015,rossi_measurement-based_2018}. As such, it allows ground state cooling~\cite{wilson_measurement-based_2015,rossi_measurement-based_2018,doherty_quantum_2012,magrini_real-time_2021,tebbenjohanns_quantum_2021} but  precludes the generation of non-classical states such as quantum squeezed states, where either the position or momentum is localised with better precision than the zero-point motion.
It has recently been predicted that the symmetry between position and momentum can be broken if the rate of the measurement is sufficiently fast~\cite{meng_mechanical_2020}. This provides the possibility of quantum squeezing, and with greatly relaxed requirements compared to other approaches\cite{brunelli2019conditional,szorkovszky2011mechanical}. 
It is known that quantum regimes can also be more easily reached when using collective interactions with multiple oscillators or non-Markovian decoherence~\cite{nielsen_multimode_2017,cripe_measurement_2019}.
However, these ideas have not yet been considered in the context of quantum state preparation by measurement.  Indeed, it might be thought that the correlations with the environment that are introduced by non-Markovian effects and the increased thermal noise from multiple oscillators would both degrade the quality of quantum state preparation.



Here we theoretically and experimentally explore measurement-based state preparation of mechanical collective modes in the presence of structural damping, an important and widespread form of non-Markovian decoherence~\cite{groblacher_observation_2015}. We show theoretically that both structural damping
and collective measurements significantly reduce the requirements to generate quantum squeezed states.
As a proof-of-principle experiment, we use optomechanical measurement to conditionally cool a structurally damped mechanical resonator. Extending to multiple mechanical modes, we show that it is possible to break the usual symmetry between position and momentum and thereby generate a thermally squeezed mechanical state. We experimentally validate that multimode conditioning allows a position variance below what is possible for measurements on a single resonator mode, and predict that room temperature quantum squeezing is within the reach of existing technology. Our work provides a path towards achieving non-classical states of macroscopic mechanical resonators at room temperature, and of exploiting these resonators in applications ranging from quantum sensing~\cite{cox_deterministic_2016,sayrin_real-time_2011} to tests of fundamental physics~\cite{minev_catch_2019,arndt2014testing, forstner_nanomechanical_2020}.

\section*{Results}

\subsection*{Optomechanical device}

A schematic of our multi-mode optomechanical experiment is shown in Fig.~\ref{Fig:Diagram}{\bf a}. It is based on a double-disk optomechanical resonator, consisting of two vertically stacked silica  microdisks, that naturally supports mechanical modes of various frequencies. Unlike previous double-disks, the structure is engineered to maximally isolate the two lowest frequency mechanical modes (the  fundamental symmetric and anti-symmetric flapping modes) from higher frequency modes. This allows measurement-based state preparation to be explored on two relatively isolated modes before considering a larger ensemble. The key features of the design, arrived at via multi-parameter finite-element optimisation (see Methods), are an asymmetric opening through the interior of each disk and a centrally-offset pedestal. Together, they result in a frequency separation that is 1.8 times larger than previous double-disk designs (\textit{e.g.}, Ref.~\cite{rosenberg_static_2009}).

Our structurally-engineered double-disk is fabricated using electron beam lithography following the process outlined in Refs.~\cite{bekker2018free, bekker2021optically}. An optical microscope image is shown in Fig.~\ref{Fig:Diagram}{\bf b}. 
The device supports hybridised optical whispering gallery modes, resulting in optical resonance frequencies that are highly sensitive to the motion of the disk edges, especially in the vertical direction~\cite{rosenberg_static_2009}. We couple an optical double-disk resonance with a decay rate of $\kappa/2\pi=2.8$~GHz to a 
fibre-based Mach-Zehnder interferometer as shown in Fig.~\ref{Fig:Diagram}{\bf c}. Performing homodyne detection returns a photocurrent that is proportional to the phase quadrature of the output optical field. The power spectral density (PSD) of this photocurrent (shown for the two lowest frequency mechanical modes in Fig.~\ref{Fig:Spectra}{\bf a}\&{\bf b}) allows us to establish the resonance frequency $\Omega_j$, decay rate $\Gamma_j$, and (boosted) optomechanical coupling rate $g_j = \sqrt{\mu_j \kappa}/2$ of each mechanical mode of the double-disk (labelled by the subscript $j$), where $\mu_j = C_j \Gamma_j$ is the mode's optomechanical measurement rate, with   $C_j$ its optomechanical cooperativity. The average thermal occupancy  of each mode $n_{\mathrm{th},j} \approx k_{\rm{B}}T/\hbar\Omega_j$ is determined by assuming that the system is in thermal equilibrium at room temperature $T$, while finite element simulations provide the effective mass $m_{\text{eff,}j}$.

\subsection*{Structural damping}

 Analysis of the observed power spectral density  allows
us to identify that the mechanical modes of our double-disk device are structurally damped. The power spectrum of thermal noise driving a resonator is given by ${S_{FF}^{\rm th}(\omega)=2 n_{\text{th}} \Omega^2 \phi(\omega)/\omega}$,
where $\omega$ is the angular frequency and $\phi(\omega)$ is the loss angle~\cite{saulson_thermal_1990, fedorov_evidence_2018,cripe_measurement_2019}.
In contrast to viscous damping, where the loss angle is linearly dependent on frequency and results in a flat thermal noise spectrum, simple models of structural damping have a frequency independent loss angle ($\phi=Q^{-1}$, where $Q=\Omega/\Gamma$ is the mechanical quality factor).
As a consequence, the thermal noise spectrum of structural damping shows a characteristic  inverse-frequency (or ``$1/f$'') dependence  \cite{fedorov_evidence_2018,cripe_measurement_2019}. 
We observe this $1/f$ dependence for the fundamental anti-symmetric flapping mode over a frequency range from $10$~kHz to $200$~kHz. As shown in Fig.~\ref{Fig:Spectra}{\bf a}, the spectrum deviates from that expected for a viscously damped oscillator by more than an order of magnitude at low frequencies, while other noise sources are 
more than 30~dB below the measured thermal noise level over the majority of the measured frequency range (see Supplementary Information). Similarly, we find that a structural damping model accurately fits interferences observed between higher frequency modes in the power spectral density, while an accurate fit is not possible using a viscously damped model.


It is known that the simple $1/f$ noise model of structural damping is not physically realistic, violating the fluctuation-dissipation theorem~\cite{saulson_thermal_1990} and causing the mechanical position spectrum to diverge at low frequencies. This results in both an infinite position variance and infinite energy, prohibiting formal estimation of the mechanical state. 
Rather than the frequency-independence of usual structural damping models, for physically realistic damping mechanisms the loss angle must be an odd function of $\omega$ to satisfy the fluctuation-dissipation theorem ~\cite{saulson_thermal_1990,landau_statistical_1980}. 
Hence, it must pass through zero at $\omega = 0$, causing a low frequency plateau in the thermomechanical noise. To our knowledge, no experimental or theoretical determination has previously been made of the frequency at which the transition to this plateau occurs, nor is the exact low-frequency functional form of $\phi$ known.

Without prior knowledge of the functional form of the loss angle, to allow state estimation we choose the simple modification $\phi={\omega}Q^{-1}/{\sqrt{\omega^{2}+\omega_{c}^{2}}}$. This is a smooth function of frequency that satisfies the fluctuation-dissipation theorem, is approximately constant above the roll-off frequency $\omega_c$, as required from observations of structural damping, and scales linearly with frequency beneath $\omega_c$ as required to enforce a low-frequency plateau in the thermomechanical noise.
Enforcing canonical thermal equilibrium ($\int^{+\infty}_{-\infty}S_{qq} d \omega=n_{\text{th}}+1/2$) allows us to constrain $\omega_c$ for this postulated form of the loss angle. Fitting the power spectrum of the fundamental anti-symmetric flapping mode, we find that the thermal occupation is within 2\% of the room temperature thermal occupancy for roll-off frequencies between 1~kHz to 10~kHz, so that $\omega_c/2\pi$ most likely lies within this range. For the state preparation in our paper, we choose $\omega_c/2\pi =10$~kHz. We note that neither the exact value of $\omega_c$ nor the detailed form of the loss angle are crucial for the main conclusions of our paper.

\subsection*{Collective modes of motion}

Our system is far in the unresolved sideband regime for all mechanical modes that we consider ($\kappa \gg \Omega_j~ \forall~j$). In this regime, the homodyne-detected optical phase quadrature is~\cite{bowen_quantum_2015}
\begin{equation}
	Y=2 \sqrt{\eta \mu^{(N)}} \, \qN{N} + \rm{measurement\; noise},
	\label{Eqn:MeasuredPhase}
\end{equation}
where $\eta$ is the measurement efficiency and $\mu^{(N)} = \sum_{j}^{N} \mu_j$ is the collective optomechanical measurement rate. 
This provides a linear readout of the collective position operator $q^{(N)}=\sum_{j}^{N} \sqrt{\mu_j} \, q_j/\sqrt{\mu^{(N)}}$, where 
$q_j$ is mode $j$'s  dimensionless position operator and is normalised such that the zero-point motion has a variance of $1/2$. The superscript ``$(N)$'' is used throughout to represent the number of conditioned mechanical modes, though we suppress it when it is clear from context.
All mechanical modes not included in $q^{(N)}$ contribute to the measurement noise, which also encompasses vacuum fluctuations, classical laser phase noise, and electronic noise arising in the photodetector and amplifiers. A collective momentum $p^{(N)}$ can be defined by analogy to $q^{(N)}$ to satisfy the canonical commutation relation, $[ q^{(N)},p^{(N)}] = \mathrm{i}$. 

%


\subsection*{Conditional state preparation and verification}

To perform conditional state preparation, we construct optimal causal Wiener filters for the position and momentum of the collective mechanical mode under study and apply them to the measurement data. A commonly used alternative is Kalman filtering~\cite{rossi_observing_2019, magrini_real-time_2021}. However, unlike Wiener filtering, this requires detailed knowledge of the temporal dynamics of the system. This is not available for structurally damped systems since they lack a widely-accepted linear time domain model~\cite{de_silva_vibration_2007}. 

We construct the causal Wiener filter for the collective position $q^{(N)}$ by separating the measured power spectral density into a signal component from the mechanical mode(s) that participate in the collective mode, and a noise component arising from all other mechanical modes, shot noise, classical phase noise, \textit{etc}., as in Eqn~(\ref{Eqn:MeasuredPhase}). The filter can then be calculated as  
\begin{equation}
	\label{eqns:WienerQ}
	\overrightarrow{H}_{q}(\omega)=\frac{1}{M_{Y}}\left[\frac{S_{q Y}}{M_{Y}^{*}}\right]_{+},
\end{equation}
where the forward (right) arrow indicates that the estimate is causal, \textit{i.e.}, it estimates the current state from data recorded at earlier times. The cross-spectral density $S_{q Y}(\omega)$ is calculated based on the fitted optomechanical parameters of the estimated mode(s), and the causal spectral factor $M_{Y}(\omega)$ is numerically generated from a fit to the entire measured power spectrum. $\left[\cdots \right]_{+}$ denotes the causal part of the contained function. An analogous procedure is used to generate the causal Wiener filter for the collective momentum, $\overrightarrow{H}_{p}(\omega)$. 


Applying the causal Wiener filters to the measured photocurrent yields causal estimates of the position and momentum. For example, the causal position estimate is  $\overrightarrow{q}^{(N)}(t)=\overrightarrow{H}^{(N)}_{q}(t)\circledast Y(t)$, where $\overrightarrow{H}^{(N)}_{q}(t)$ is the collective position filter and  $\circledast$ denotes a convolution. To quantify the uncertainty in the estimates it is necessary to calculate the variances of their deviation from the true values. For example, the conditional position variance is given by $V_{\delta\pre{q}\delta\pre{q}} = \expect{(q^{(N)}-\pre{q}^{(N)})^{2}}$. Unlike the estimates of the position and momentum themselves, the variances cannot be directly obtained because the `true' position $q^{(N)}$ and momentum $q^{(N)}$ are experimentally inaccessible. Instead, the prepared conditional state must be verified via comparison to a second set of independent estimates.
We use
anti-causal filters ($\overleftarrow{H}_{q}$ and $\overleftarrow{H}_{p}$), which estimate the current position and momentum from measurement data at later (future) times, i.e. they perform retrodiction. These are labelled by a backward (left) arrow. The anti-causal estimate of position, for example, is calculated as
$\overleftarrow{q}^{(N)}(t)=\overleftarrow{H}^{(N)}_{q}(t)\circledast Y(t)$.

Following Rossi \textit{et al.}~\cite{rossi_observing_2019}, we define the \textit{relative position and momentum estimates}, $\Delta q(t) = \overrightarrow{q}^{(N)}- \overleftarrow{q}^{(N)}$ and $\Delta p = \overrightarrow{p}^{(N)}- \overleftarrow{p}^{(N)}$. The conditional variances have generally been thought to be related to the variances of these relative estimates by a simple factor of two~\cite{rossi_observing_2019}. However, we find that this simple relationship does not hold either for non-Markovian dynamics or for collective measurements on multiple resonators (see Methods). In general, the variances can be related by
${V_{\Delta q\Delta q} = 2 (1-F_q)V_{\delta\pre{ q}\delta\pre{ q}} }$ and ${V_{\Delta p\Delta p} = 2 (1-F_p)V_{\delta\pre{ p}\delta\pre{ p}} }$, where the conversion factors $F_q$ and $F_p$ characterise the deviation from the rotating wave result.
We use simulations to determine these conversion factors for each collective mode studied. This allows the conditional variances to be experimentally quantified.

We refer to the reader to the Supplementary Information for details on our prediction and retrodiction procedures.


\subsection*{Single mode estimation}

We begin by optimally estimating the position  $q^{(1)}$ and momentum $p^{(1)}$ of only the (lowest frequency) fundamental anti-symmetric flapping mode, with all other mechanical modes treated as noise. The magnitude response of the optimal causal Wiener filter for $\qN{1}$ is shown by the blue trace in Fig.~\ref{Fig:Spectra}{\bf c}.
Wiener filters act to accept frequency components with high signal-to-noise, and to reject those with low signal-to-noise. This is illustrated by the deep notch around the frequency of the fundamental symmetric flapping mode.
To show the effect of the Wiener filter in the time domain, we compare the measured optical phase quadrature as a function of time before and after filtering.
The left side of Fig.~\ref{Fig:Spectra}{\bf d} shows the raw (normalised) phase quadrature $Y$. The thermally-driven fundamental anti-symmetric flapping mode is the dominant mechanical contribution, due to its strong optomechanical coupling, but its motion is partially obscured by measurement noise and the dynamics of the other mechanical modes. On the other hand, the pink trace on the right side of Fig.~\ref{Fig:Spectra}{\bf d} shows the optimal causal estimate of the position $\overrightarrow{q}^{(1)}(t)$ as a function of time, calculated by applying the filter as described above
($\overrightarrow{q}^{(1)}(t)=\overrightarrow{H}^{(1)}_{q}(t)\circledast Y(t)$)
As can be seen, applying the filter removes much of the noise present in the unfiltered trace. The dark-red trace shows the anti-causally estimated position. This agrees closely with the causal estimate, as quantified by the relative estimate (blue trace).



Fig.~\ref{Fig:Spectra}{\bf e}  plots the relative estimates in mechanical phase space, together with the direct position and momentum estimates. This allows the effect of single-mode conditioning on the mechanical state to be directly visualised. The phase space distribution of the direct estimates is consistent with a room-temperature thermal state, as expected. The relative estimates, magnified in the phase space distribution of  Fig~\ref{Fig:Spectra}{\bf f}, are confined much more closely near the origin. This is an example of ``cooling by measurement''~\cite{vanner_cooling-by-measurement_2013}. 

After determining the conversion factors $F_q^{(1)}=0.26$ and ${F_p^{(1)}=-0.11}$, we find that the position and momentum conditional variances are $V_{\delta \pre{q} \delta \pre{q}}^{(1)}=2.5\times10^{5}$ and  $V_{\delta \pre{p}\delta \pre{p}}^{(1)}=2.7\times10^{5}$, respectively,  agreeing with simulations to within 6\%. These variances are reduced by factors of  $9.7\times 10^{-3}$ and $10 \times10^{-3}$, respectively, compared to the thermal variance of the mode, showing substantial conditional cooling. An upper bound on the conditional phonon occupancy of the mode can be calculated as $\bar{n}_{\text{cond}}<(V_{\delta \pre{q}\delta \pre{q}}^{(1)}+V_{\delta \pre{p}\delta \pre{p}}^{(1)}-1)/2 = 10 \times10^{-3} n_{\text{th}}$. Thus, the fundamental antisymmetric flapping mode is conditionally cooled by at least two orders of magnitude.

It is interesting to note that, while the phase space distributions of Fig.~\ref{Fig:Spectra}{\bf e}\&{\bf f} appear to show significant thermomechanical squeezing, this is in part an artifact of the differences between the conversion factors for position and momentum. The effect remains once the relative variances are converted into conditional variances, showing that our measurements are outside of the regime of validity of the rotating wave approximation. However, it is greatly reduced -- the ratio of momentum to position variances is 1.7 for the relative variances, but only 1.1 for the conditional variances. 
This highlights the importance of correct accounting of the relationship between variances.

The relatively low level of thermomechanical squeezing also highlights the detrimental effects of the higher order mechanical modes on the estimation process.  For an isolated viscously damped resonator, Ref.~\cite{meng_mechanical_2020} derived the criterion $S = 16 \mu n_{\text{tot}}\Gamma/\Omega^2 > 1$ for thermal squeezing, where $n_{\text{tot}}= n_{\text{th}} + C + 1/2$ is the total effective occupancy of the mode. For our experiments, the fundamental antisymmetric flapping mode has $S=3.5\times10^3 \gg 1$.  Indeed, including structural damping, our simulations show  that, were the fundamental antisymmetric flapping mode well isolated from all other modes, the conditional state should be extremely strongly squeezed, having a ratio of momentum to position variances of 60. The problem of the presence of additional mechanical modes can be avoided, for example, by using a single-mode pendulum system, as has recently been used to demonstrate thermomechanical squeezing by Matsumoto {\it et al.}~\cite{matsumoto_preparing_2020}.  As we will show  in what follows, multimode conditioning provides an alternative and more generally applicable solution.

\subsection*{Multimode estimation}

Multimode conditioning achieves two closely-related goals; it reclassifies the higher-order modes from noise to signal, and in the process increases the effective measurement strength~\cite{xuereb_collectively_2013,kipf_superradiance_2014,nair_cavity_2016,piergentili_two-membrane_2018,xuereb_strong_2012}. We first examine the two-mode case, estimating the collective position and momenta of the well frequency-separated fundamental symmetric and anti-symmetric flapping modes. Including these two modes increases the optomechanical measurement rate from $\mu^{(1)}/2\pi=0.86$~kHz to $\mu^{(2)}/2\pi=0.91$~kHz.
The causal Wiener filter for the collective position,  calculated via Eq.~(\ref{eqns:WienerQ}), is shown in green in Fig.~\ref{Fig:Spectra}{\bf c}. Applying this filter and its anti-causal counterpart, we find that the collective conditional position variance of this two-mode collective mode is $V_{\delta \pre{q}\delta \pre{q}}^{(2)}=1.6\times10^{5}$ with $F_q^{(2)}=0.39$. Similarly, we find a collective momentum variance of $V_{\delta \pre{p}\delta \pre{p}}^{(2)}=2.0\times10^{5}$ with $F_p^{(2)}=-0.18$. These variances are respectively $6.4 \times10^{-3}$ and $7.9 \times10^{-3}$ times smaller than the two-mode thermal variance, and are improved by factors of 1.6 and 1.4, respectively, compared to single mode estimation. Since the two-mode estimate reduces the variance in (collective) position more than in momentum, it increases the thermomechanical squeezing. This results in a ratio of momentum to position variances of 1.2. A similar improvement in squeezing of the relative estimates is seen in phase space (green dots) in Fig.~2{\bf e}\&{\bf f}.

 
Our two-mode estimate, while superior to a single mode estimate, is still roughly a factor of 400 inferior to predictions from simulations with no other mechanical modes present in the detected photocurrent.  The state preparation can be improved further by including a larger number of mechanical modes in the collective mode. As an example,  
we select the first five modes, shown with red shading in the power spectral density in Fig.~\ref{Fig:Multimode}{\bf a}. The optimal causal Wiener filter for the position of this five-mode collective mode is shown in Fig.~\ref{Fig:Multimode}{\bf b}. It exhibits a broadband response, with slow modulations due to changes in the signal-to-noise ratio and several sharp notches which arise due to noise peaks in the power spectral density (shaded grey in Fig.~\ref{Fig:Multimode}{\bf a}). Applying this filter, the analogous filter for momentum, and the counterpart anti-causal filters results in the experimental relative conditional state visualised in Fig.~\ref{Fig:Multimode}{\bf c}. 
As can be seen, the five-mode collective state 
breaks the RWA more strongly than the one- and two-mode states, being significantly more elliptical in phase space.  This can be expected since it has a collective measurement rate of $\mu^{(5)}/2\pi=1.0$~kHz, higher than the measurement rates for either single or two-mode cases.  
We calculate the conditional position variance to be $V_{\delta\pre{q} \delta\pre{q}}^{(5)} = 1.1\times10^5$, $5.0\times10^{-3}$ times smaller than the five-mode thermal variance, with $F_q^{(5)}=0.21$. The ratio of momentum to position conditional variances  is found to be 2.3, almost a factor of two larger than the two-mode case. Increasing the number of modes to nine further improves the measurement rate and thermal squeezing ratio to $\mu^{(9)}/2\pi=1.3$~kHz and 2.7, respectively (See Supplementary Information). 




The better conditioning we achieve using collective mechanical modes implies that the measurement is inducing conditional correlations between the mechanical modes. 
%
We show these correlations experimentally, choosing to examine position-position and momentum-momentum correlations between two collective modes. The collective modes we choose are the two-mode collective mode comprising the lowest and second-lowest frequency modes (position, $\qN{2}$; momentum, $p^{(2)}$), and the three-mode collective mode comprised of the third, fourth and fifth lowest frequency modes, with collective position and momentum denoted here as $\qN{*3}$ and   $p^{(*3)}$, respectively. Fig.~\ref{Fig:Multimode}{\bf d} plots the relative position estimates of the two collective modes against each other, while Fig.~\ref{Fig:Multimode}{\bf e} plots the relative momentum estimates.
A negative correlation of ${\langle \qN{2} \qN{*3} \rangle/\sqrt{\langle \left. q^{(2)\,}\right.^2 \rangle \langle \left. q^{(*3)\,}\right.^2\rangle} = -0.38}$ is observed between the relative collective positions; while a positive correlation of $+0.09$ is observed between momenta, calculated in the same way. 
The fact that the observed correlations are of opposite sign suggests that stronger measurements could generate conditional entanglement. 
In the Methods we show that this is indeed possible for viscous damping, deriving the entanglement condition ${C>n_{\text{th}}^2 Q/2N}$.


\subsection*{Quantum squeezing of structurally damped oscillators}

In the above sections, we have demonstrated measurement-based state preparation for the first time for a structurally damped resonator, and have shown that multi-mode conditioning can improve the preparation of classical thermomechanical squeezed states. It is interesting to ask how structural damping and multi-mode conditioning affect the ability to generate quantum squeezing. 
It is known that measurements of the collective  motion of multiple mechanical modes yield a larger effective measurement strength than single mode measurements~\cite{xuereb_collectively_2013,kipf_superradiance_2014,nair_cavity_2016,piergentili_two-membrane_2018,xuereb_strong_2012,nielsen_multimode_2017}. However, to our knowledge, whether this can allow more effective ground state cooling or quantum squeezing has not previously been addressed.  Similarly, by reducing the high frequency thermal noise of a mechanical resonator, structural damping has been shown to relax the requirements to observe radiation pressure backaction~\cite{cripe_measurement_2019} and to achieve quantum squeezing of light~\cite{aggarwal_room-temperature_2020}. However, one might expect that the increased thermomechanical noise at low frequencies would degrade the ability to prepare mechanical non-classical states.

To address these questions, we theoretically model an optomechanical measurement of a collective mode of a structurally damped resonator (see Methods). We compare the results to the criterion for quantum squeezing of a single viscously damped resonator derived in Ref.~~\cite{meng_mechanical_2020}. For simplicity, we consider only the case where all $N$ modes are identical and experience the same measurement rate $\mu_j$ so that the collective measurement rate $\mu^{(N)}=N \mu_j$.
The case of multiple mechanical modes with varying properties is substantially more complex to treat theoretically and left for future work. We find analytically that the use of multiple modes relaxes the required cooperativity to achieve quantum squeezing by a factor of $1/N$, a result that we show also holds for a viscously damped oscillator. Through simulations, we find that structural damping also reduces the cooperativity needed for quantum squeezing, in this case by a factor of $\left(Q/n_{\mathrm{th}}\right)^{1/12}$. Overall, this results in the condition $C > n_{\text{tot}}^{(N) 1/4}Q^{3/4}/N$ for quantum squeezing of a structurally damped multi-identical-mode mechanical system, where ${n_{\mathrm{tot}}^{(N)}=n_{\mathrm{th}}+NC+1/2}$ is the total thermal occupancy of the common mechanical mode
, with $n_{\mathrm{th}}$ being the occupancy in the absence of measurement-induced heating. Note that the optical heating term is boosted by $N$ because every mechanical oscillator experiences the same backaction force, \textit{i.e.}, the optical noise is perfectly correlated (the thermal forces, by contrast, are uncorrelated). 

To examine the experimental feasibility of achieving quantum squeezing with a structurally damped mechanical resonator, we consider a structurally damped zipper cavity device of the form reported by Leijssen {\it et al.} in Ref.~\cite{leijssen_nonlinear_2017}. This device had a mechanical frequency $\Omega/2\pi$=3~MHz, a
single photon optomechanical coupling strength of $g_0 /2\pi$=24~MHz, and mechanical and optical decay rates of $\Gamma/2\pi=100$~Hz and $\kappa/2\pi=20$~GHz, respectively. For a single mechanical resonance and assuming a detection efficiency of $\eta=0.5$, these parameters should allow conditioning quantum squeezed states from room temperature with only 160 intracavity photons. Were an array of identical resonators used, the required intracavity photon number would drop even further.

\section*{Discussion}

To conclude, we have demonstrated the preparation of mechanical states by measurement, beyond previous work focused on single viscously damped resonators~\cite{rossi_observing_2019,matsumoto_preparing_2020, brawley_nonlinear_2016}. We have shown experimentally that state conditioning and verification is possible for structurally damped resonators, and that collective measurements on multiple mechanical modes can improve state preparation. Our theoretical modelling shows that structural damping can, somewhat remarkably, improve the performance of state preparation, as can the use of multiple mechanical modes. This substantially reduces the requirements to reach quantum regimes, with room temperature quantum squeezing within the reach of existing technology. As such, our results open a pathway to room temperature quantum optomechanical technologies, such as quantum force sensors and interfaces, and to explore the crossover between quantum and classical physics.

A significant technical advance in our paper is the development of a method to determine the roll-off frequency of structural damping.
Structural damping is encountered in many macroscopic engineered structures \cite{kimball_internal_1927,saulson_thermal_1990,de_silva_vibration_2007} and has been observed in a range of optomechanical systems~\cite{neben_structural_2012,safavi-naeini_squeezed_2013,fedorov_evidence_2018, cripe_measurement_2019,komori_attonewton-meter_2020}. However, there is no widely accepted model for its origins~\cite{groblacher_observation_2015}. A low frequency roll-off must exist to satisfy thermodynamics, but to our knowledge has never been observed in experiments. Our use of the equipartition theorem to constrain the value of the cut-off frequency, even when it is beneath experimentally resolvable frequencies, provides a means to access new information about structural damping. Moreover, the kilohertz-range roll-off frequency indicated by our analysis suggests that direct experimental measurements of the low-frequency behaviour of structural damping may be within reach. Together, these possibilities provide a new route towards understanding the underlying mechanisms that give rise to this widespread and relatively poorly understood form of damping~\cite{groblacher_observation_2015}.

\newpage

\begin{figure}[H]
	\centering
	\begin{subfigure}{0.5\textwidth}	\includegraphics[width=\textwidth]{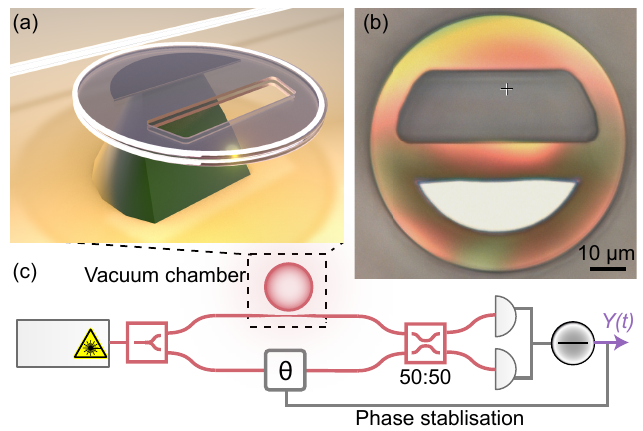}
	\end{subfigure}
	\caption{\label{Fig:Diagram} {\bf Experimental apparatus.} (a)~Illustration of the structurally engineered double-disk optomechanical resonator and tapered fibre used for optical excitation. (b)~Microscope image of the optomechanical resonator. The key dimensions of the resonator are: diameter, $70$~$\upmu$m; thickness, $400$~nm; and separation between the disks, $\sim 300$~nm. (c)~Schematic of  homodyne measurement of optomechanical system. The laser is tuned onto resonance with the optomechanical resonator. The optomechanical device is housed in a vacuum chamber (pressure $<10^{-6}$~mbar), as indicated by the black dashed box. The photocurrent $Y(t)$ is digitised on an oscilloscope and stored for post-processing.}
\end{figure}

\begin{figure}[H]
	\centering
		\includegraphics[width=\textwidth]{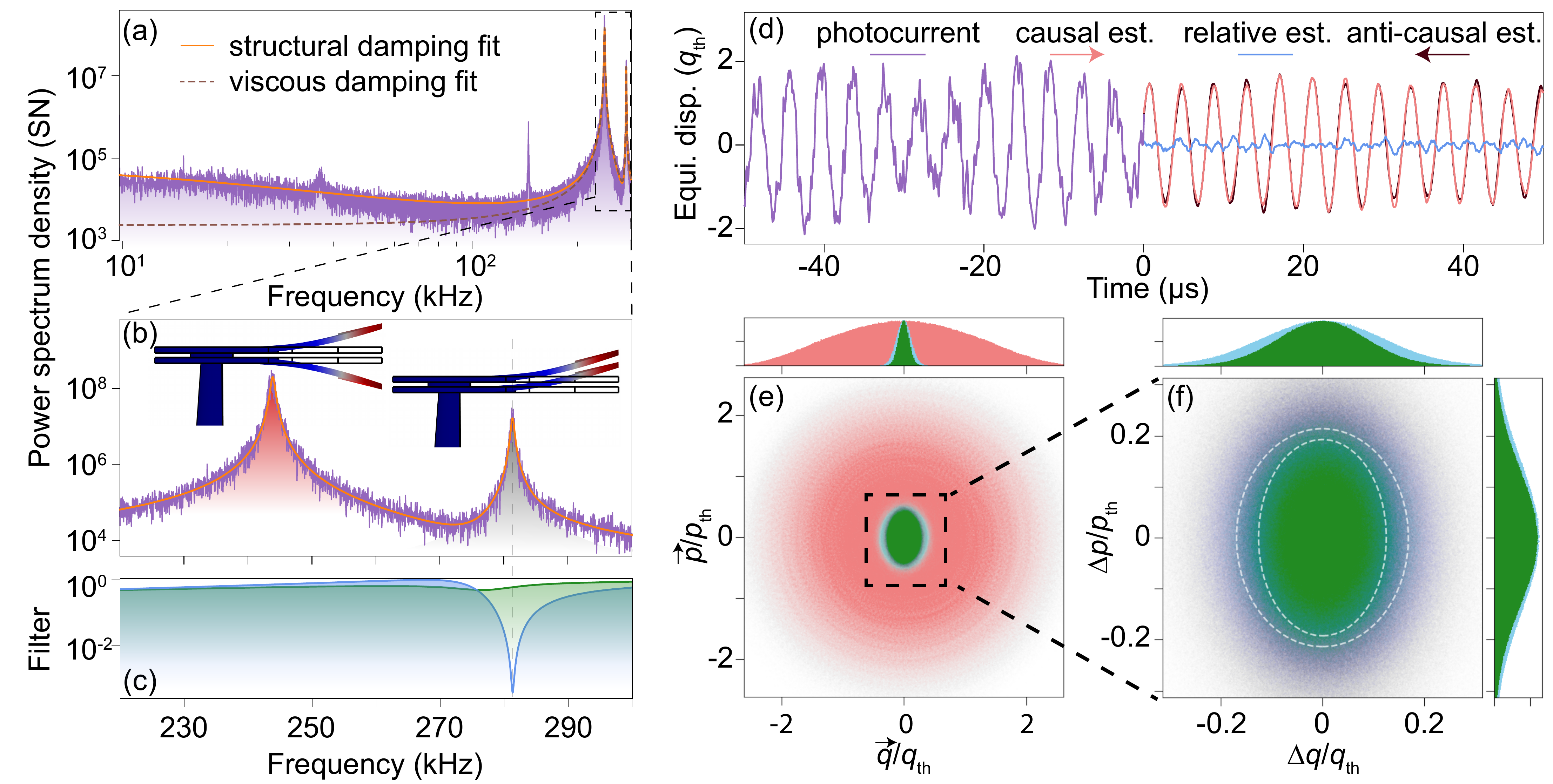}
	\caption{\label{Fig:Spectra} {\bf Preparation of single- and two-mode mechanical states of a structurally damped resonator.}  (a)~Power spectral density of thermal noise (purple), and fitted curves using structural damping (orange) and viscous damping (brown dashed line) , normalised such that the shot noise level is $1/2$. (b)~Zoom-in of the thermal noise power spectral density (purple) of the first two mechanical modes. Red (grey) shading indicates the fundamental anti-symmetric (symmetric) flapping mode. These modes are calibrated to have  $(\mu_1, \Omega_1, \Gamma_1)/2\pi=(0.74, 244, 0.69)$~kHz and $(\mu_2, \Omega_2, \Gamma_2)/2\pi=(0.053, 281, 0.50)$~kHz, respectively.
    (c)~Magnitudes of filter functions. Blue: single-mode estimation; green: two-mode estimation.
    (d)~Time traces of the photocurrent (purple), the causal (pink) and anti-causal (maroon) position estimates of the first mechanical mode, and the corresponding relative estimate (blue), normalised by the standard deviation of the thermal fluctuations, ${q_{\text{th}}=(\int^{\infty}_{-\infty} S_{qq}(\omega){d\omega}/{2\pi})^{1/2}}$. 
	(e)~Phase space distributions of the unconditional state (pink), and relative single-and two-mode estimated states (blue and green, respectively). Axes normalised to the standard deviation of the thermal position and momentum (${p_{\text{th}}=(\int^{\infty}_{-\infty} S_{pp}(\omega){d\omega}/{2\pi})^{1/2}}$) fluctuations.   (f)~Magnified phase space distributions of the one- and two-mode relative estimates. White dashed ellipses show their respective Wigner function contours. (e) and (f) include histograms (4 million data points) of the position and momentum distributions, which obey Gaussian statistics. }
\end{figure}


\begin{figure}[H]
	\centering
	\includegraphics[width=\textwidth]{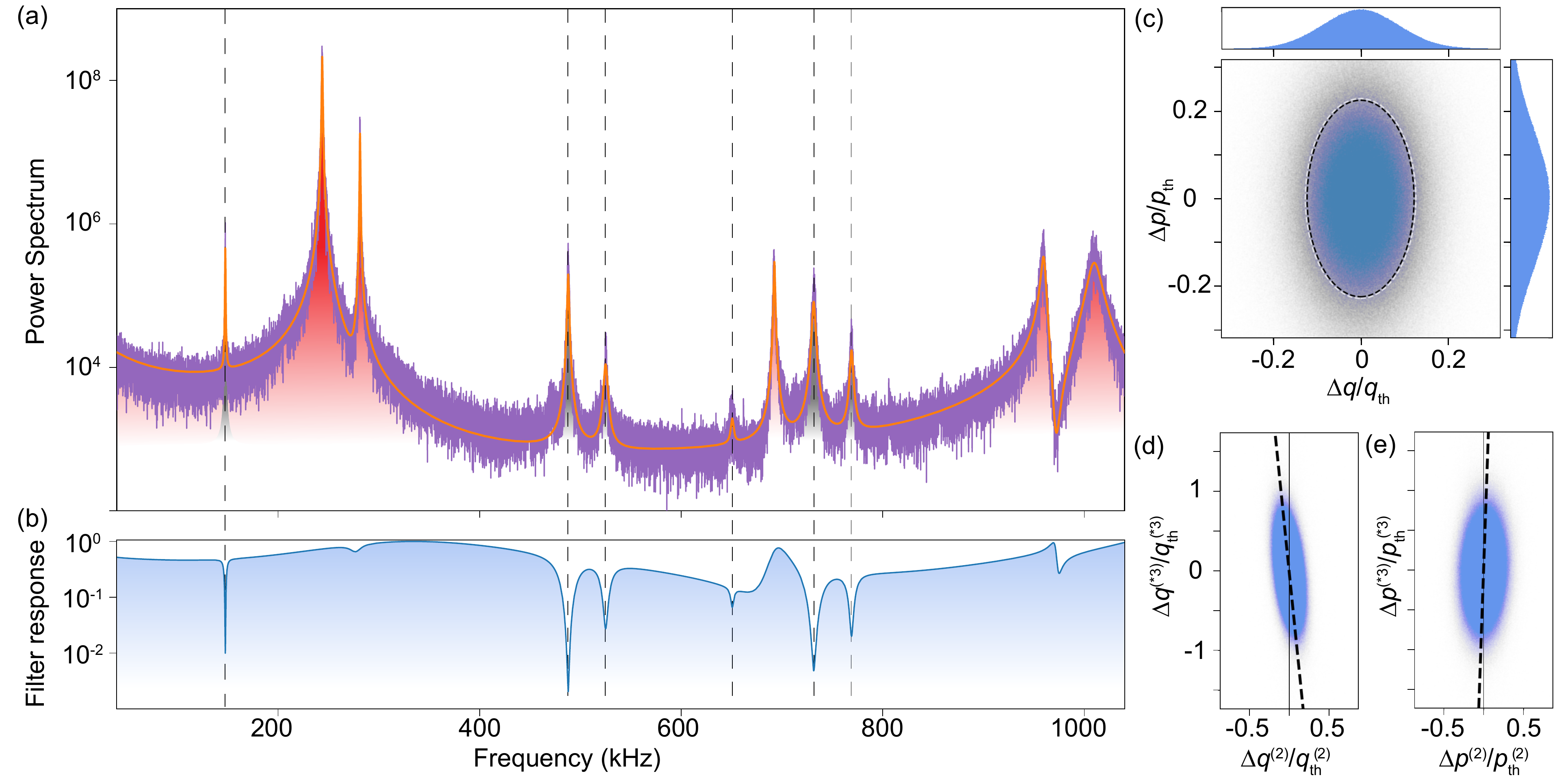}
	\caption{\label{Fig:Multimode} {\bf Preparation of five-mode mechanical states}. (a)~Power spectral density of thermal noise (purple), and fitted curve using structurally damped model (orange). The red shaded areas are mechanical peaks and gray ones are noise peaks including non-mechanical and nonlinear peaks. The lowest frequency of these noise peaks is not of mechanical origin modelled with Lorentzian power spectra, and the other five originate from nonlinearities in the optomechanical interaction modelled with structural damping power spectra. They are all treated as noise. (b)~Filter response for conditioning the five mechanical modes. The filter function response notches out all the noise peaks (indicated by the vertical dashed lines). (c)~Phase space distribution of the relative estimates of the five mechanical modes. Black and white dashed circles show the contour of the Wigner function from simulation and experiment, respectively, with very close agreement. (d)~\&~(e)~Position and momentum correlations between collective modes made up of the lowest and second lowest frequency mechanical modes, and the third, fourth and fifth lowest frequency modes, with normalised correlations of $-0.38$ and $0.09$ respectively.}
\end{figure}


\section*{Methods}
\subsection*{Optomechanical system and measurement}

Our structurally-engineered double-disks have been optimised to maximise the frequency separation of adjacent mechanical modes; this is achieved by incorporating off-centre slots that result in asymmetric pinning, a key characteristic that differs from previous designs~\cite{rosenberg_static_2009, bekker2018free, bekker2021optically}. Qualitatively speaking, this forces the mechanical modes to behave more like those of a singly-clamped cantilever, which are widely separated in frequency. Specifically, typical symmetrically-clamped double-disk resonators exhibit a pair of low frequency modes, with the next mode occurring at a frequency around 1.6 times the frequency of the fundamental (lowest frequency) mode~\cite{rosenberg_applications_2010}. By comparison, in our asymmetrically-clamped double-disks this third mode is shifted up to 2.9 times the fundamental frequency. This increased separation is in close alignment to the case for a singly clamped cantilever oscillator, for which the third mode is at approximately three times the fundamental frequency. 

The asymmetric pinning in our structurally-engineered double-disk design acts to lengthen the mechanical arms of the oscillator, which lowers the frequencies of the fundamental mechanical modes $\Omega_1$ and $\Omega_2$. This assists in the preparation of squeezed states, as discussed in Ref.~\cite{meng_mechanical_2020}.
The slots in our double-disk also act to reduce the effective mass ($m_{\text{eff,1}}$=536~pg for the  fundamental anti-symmetric flapping mode).  As a result, our design increases the zero-point motion of fundamental mode by a factor of 2.1 compared to a spiderweb double-disk~\cite{rosenberg_static_2009} and 5.7 compared to a solid double-disk~\cite{lin_mechanical_2009,rosenberg_applications_2010}. Counteracting these improvements, the asymmetric pinning suppresses motion on one half of the disk, reducing the optomechanical coupling strength by a factor of 3.6 to \mbox{4.3~GHz/nm}, a reduction which is consistent with our experimental observations~\cite{rosenberg_applications_2010}. Overall, by permitting the thermal noise of the two lowest frequency modes to dominate the power spectral density over a larger bandwidth than a comparable symmetrical double-disk, our design allows better estimation of the motion of the collective mode that consists of the lowest and second lowest frequency modes.

A continuous-wave diode laser running at the wavelength of 1555~nm is resonant with the whispering-gallery mode of the double-disk device. For a fixed optical power injected into the optomechanical system, it is desirable to work near the critical coupling point, where the intrinsic cavity energy damping rate $\kappa_{0}$ is equal to the input--output coupling rate $\kappa_{\text{ex}}$, since this maximises the effective measurement rate ${\mu \propto \kappa_{\text{ex}}^2/(\kappa_0+\kappa_{\text{ex}})^4}$. In practice, we couple the light in and out by positioning a tapered optical fibre close to the perimeter of the double-disk.  Positioning limitations mean that we operate in the slightly-overcoupled regime, with $\kappa_0/2\pi=1.0$~GHz and $\kappa_{\text{ex}}/2\pi=1.8$~GHz. Injecting 3.7~$\upmu$W of optical power results in $4300$ intra-cavity photons. The total detection efficiency is $\eta=30\%$ (including the escape efficiency of 64\%, fibre transmission 67\%, taper transmission 89\%, and detector quantum efficiency 80\%).

\subsection*{Double-disk design optimisation}

\begin{figure}[H]
	\centering
	\includegraphics[width=0.7\textwidth]{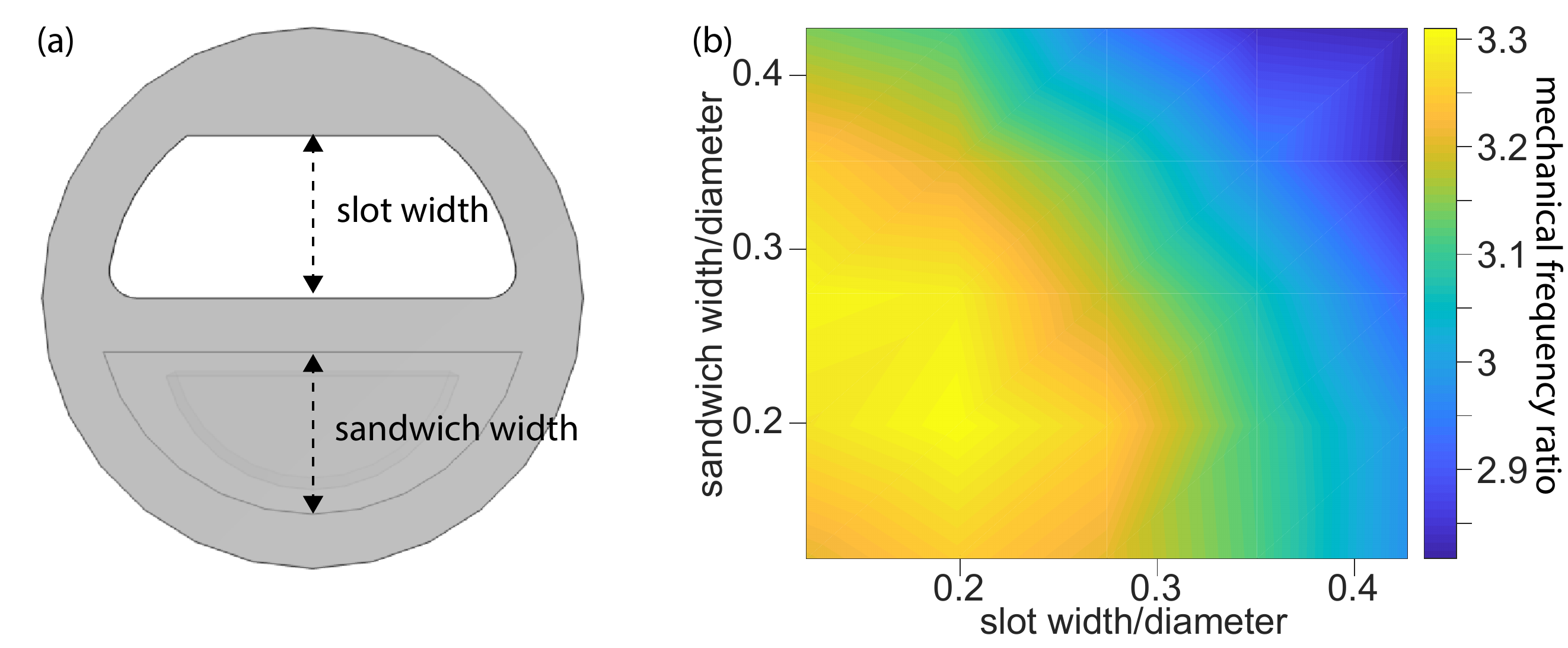}
	\caption{\label{Fig:Engineered_DD} {\bf Optimisation of the structure of the double-disk.} (a) Diagram of the engineered structure. The width of the slot and sandwich layers are labelled with black dashed lines. The half circle inside the sandwich layer represents the top of the pedestal which supports the double-disk. This is narrower than the sandwich layer due to the faster etch rate of the pedestal.   (b) Colormap of the ratio of the frequencies of the second to first mechanical eigenmode as a function of the slot and sandwich layer widths, normalised to the diameter of the disk.}
\end{figure}

The geometry of the double-disk structure is optimised to maximise the ratio of the frequencies of the high order modes to the fundamental flapping modes. As the symmetric and antisymmetic fundamental modes are close to each other in frequency, and because our fabrication technique leads to roughly identical slot shape and size on both top abd bottom disks,  we simplified the model by simulating one disk with one slot. For a fixed diameter of disk, the mechanical frequencies are affected by the location and geometry of the slot and of the sandwich layer between the disks (which acts as a clamping point for the mechanical motion). We choose the slot geometry to be a half-circle with a flattened top, as shown in Fig.~\ref{Fig:Engineered_DD}{\bf a}. The flattened top reduces the width of the slot, and increases the effective mass of the fundamental mode without significantly changing its spring constant. This decreases its frequency. On the other hand, it can stiffen the higher frequency modes, and has a smaller effect on their effective masses. Together, this results in an increased frequency separation compared to alternative double-disk designs.

The slot geometry is defined by the mask used in electron beam lithography. The radius of the circular part is chosen to ensure that the rim of the disk is wide enough to contain the whispering gallery modes of the device, and therefore that the slot does not introduce significant optical losses. In fabrication, the etchant reaches the sandwich layer both from outside the disk and by diffusing through the slot. This results in the semi-circular sandwich layer geometry shown in Fig.~\ref{Fig:Engineered_DD}{\bf a}. By tuning the duration of the etch we are able to tune the width of the sandwich layer. 

Given that the slot and sandwich layer widths can both be tuned in fabrication, we perform finite-element simulations to investigate what the optimum choice of these two parameters is to maximally separate the frequency of the fundamental mode from the higher frequency modes. Specifically, we determine the ratio of the frequency of the lowest frequency mode to that of the second lowest mode, as as function of slot and sandwich layer widths. We note that including two disks, rather than the one modelled here, the lowest frequency mode separates into the symmetric and antisymmetic fundamental modes we observe in experiment. The results of the simulation are shown in Fig.~\ref{Fig:Engineered_DD}{\bf b}, where the sandwich and slot widths are normalised to the diameter of the disk.  
As can be seen, the maximum modes separation of around a factor of three occurs when the slot width and sandwich width are both around 20\% of the disk diameter.  

\subsection*{Model of multimode state preparation with viscous damping}


To understand theoretically how collective measurements on multiple modes differ from the single mode measurements considered by Meng {\it et al.}~\cite{meng_mechanical_2020}, we restrict ourselves to the simple case where the mechanical modes are identical (equal $C$, $\Omega$, $\Gamma$, \textit{etc.}). It is then possible to directly use the results of Meng \textit{et al.}~\cite{meng_mechanical_2020} to predict the effectiveness of localisation of a viscously damped collective mode; one need only make the simple replacement $C\rightarrow NC$. We find that the rotating wave approximation breaks down when $C > Q^2/N \eta n_{\mathrm{tot}}^{(N)}$. 
The breakdown of the rotating wave approximation, and therefore generation of thermal squeezing, occurs at cooperativities a factor of $N$ lower than those found in Meng \textit{et al.}~\cite{meng_mechanical_2020}. Moreover, we find that -- deep within the regime where the rotating wave approximation breaks down -- the conditional variance of the collective position becomes
\begin{equation}
\label{Eqns:Common_position_variance}
	V_{\delta q \delta q}^{(N)}\approx \left [  \frac{Q^2 n_{\mathrm{tot}}^{(N)}}{64(\eta N C)^{3}} \right ]^{1/4}.
\end{equation}
Compared to the single mode case, this variance is reduced by a factor of $N^{3/4}$ in the thermal noise dominated regime and $N^{1/2}$ in the backaction dominated regime.

The purity of the conditional state of ${\mathcal{P}=\sqrt{{\eta N C}/{n_{\text{tot}}}}}$ is increased, but saturates to the same value in the  backaction dominated regime.  We see therefore that collective measurements on multiple mechanical oscillators can both greatly relax the requirement to prepare a non-classical state, and improve the quality of the non-classical state that is prepared. 

Within the regime of validity of the rotating wave approximation, collective measurements on multiple mechanical modes can also improve cooling by measurement. Specifically, the condition for when the ground state of a viscously damped resonator is approached is relaxed by a factor of $N$ to $C > n_{\rm th}/N$.

\subsection*{Model of state preparation with structured damping}

\begin{figure}[t]
	\centering
		\includegraphics[width=0.5\textwidth]{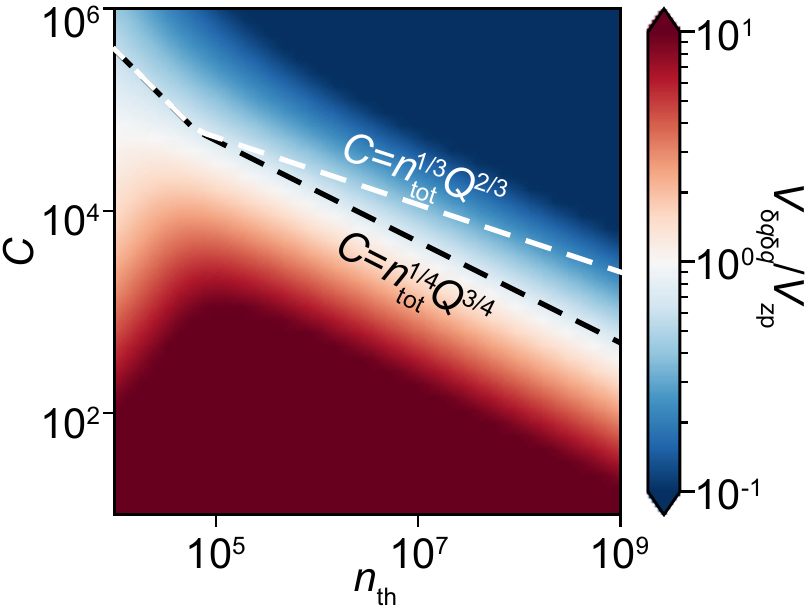}
		\caption{\label{Fig:Structural damping theory}Simulation of the conditional position variance of a single-mode structurally damped resonator under continuous position measurement. The state becomes conditionally quantum squeezed above the black dashed line. The threshold for quantum squeezing of a single-mode viscously damped resonator is shown for comparison by the white dashed line. We see that structural damping results in relaxed experimental requirements to achieve quantum squeezing in the regime where $C\ll n_{\text{th}}$.}
\end{figure}

For structurally damped resonators, we begin by modelling a single-mode mechanical resonator and extend to the multimode case in the same way as in the viscous damping case treated above.
To make qualitative conclusions we consider continuous position measurement and
numerically calculate the conditional position variance as a function of $C$ and $n_{\text{th}}$ (assuming unity detection efficiency). The results are shown in Fig.~\ref{Fig:Structural damping theory}. Here we fix the bath at room temperature so that an increase in $n_{\text{th}}$ corresponds to a decrease in the mechanical frequency. We observe that quantum squeezing can be achieved when $C>n_{\text{tot}}^{1/4} Q^{3/4}$ (dashed black line in Fig.~\ref{Fig:Structural damping theory}), \textit{i.e.} with lower $C$ compared to a single-mode viscously damped oscillator in the thermal noise dominated regime~\cite{meng_mechanical_2020}  (dashed white line in Fig.~\ref{Fig:Structural damping theory}). For $N$ identical structurally damped modes, the quantum squeezing criterion is relaxed to $C>n_{\text{tot}}^{(N)1/4} Q^{3/4}/N$ by replacing $C\rightarrow NC$, as in the viscous damping case.

\subsection*{Relationship between relative and conditional variances for a structurally damped resonator}

\begin{figure}[H]
	\centering
		\includegraphics[width=0.7\textwidth]{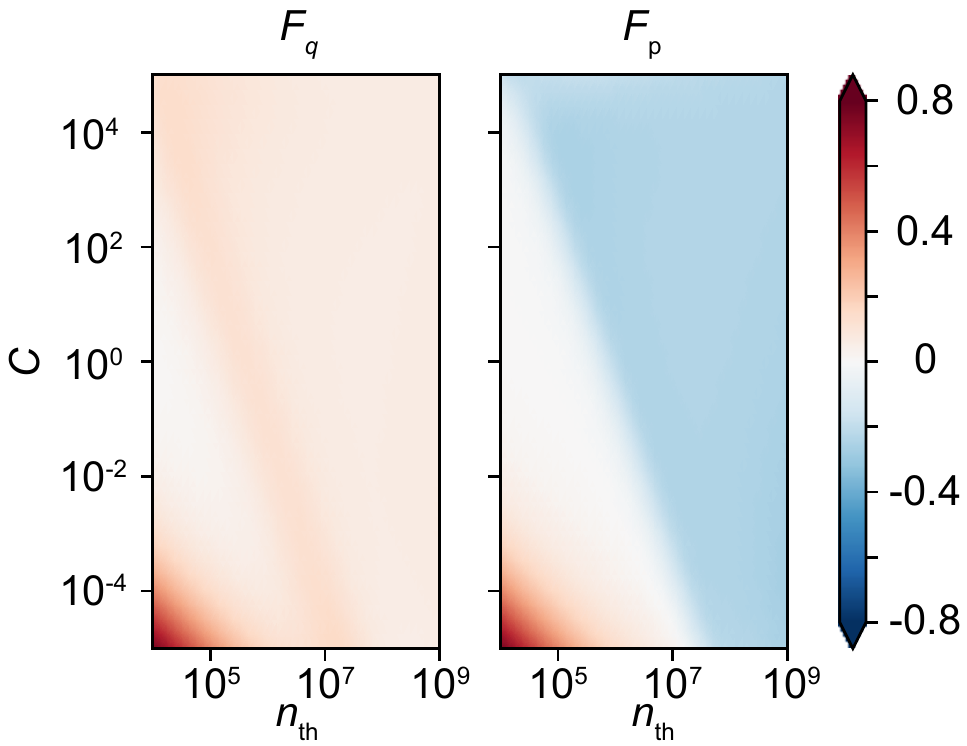}
		\caption{\label{Fig:Fqp} Characterisation of the conversion factors $F_q$ and $F_p$ for a single-mode structurally damped oscillator as a function of $n_{\text{th}}$ (fixing the bath at room temperature) and $C$. (I will fix font and style later.)}
\end{figure}

In our experiments the conditional state is prepared and verified using prediction and retrodiction. In the case of a single viscously damped resonator and in the limit of a high measurement rate, the conditional variance is equal to half the variance of the difference between the estimates produced via prediction and retrodiction (i.e. the variance of the relative estimate)~\cite{rossi_observing_2019}. However, we observe that this is not the case for a structurally damped oscillator even in the high measurement regime. As a result, characterisation of the conversion factors $F_q$ and $F_p$ is essential to obtain the conditional state. We characterise them via simulations. As an example, Fig.~\ref{Fig:Fqp} shows both conversion factors as a function of the thermal occupancy and cooperativity. As can be seen, $F_p$ becomes negative at sufficiently high cooperativities, with the regime over which this occurs roughly coinciding with the regime where the rotating wave approximation is invalid. $F_q$, on the other hand is always positive but is non-zero at high cooperativities. 

\subsection*{Analysis of measurement-induced entanglement}

In our discussion of measurement-induced entanglement in the main text, we consider a specific form of entanglement in which an even  number $N$ of identical mechanical modes (equal $C$, $\Omega$, $\Gamma$, \textit{etc.}) are measured. The common position ($q_1+q_2+q_3+q_4...q_{N-1}+q_{N}$) is strongly conditioned (as discussed above). Conversely, the differential momentum ($p_1-p_2+p_3-p_4...+p_{N-1}-p_{N}$) is entirely decoupled from the dynamics of the common position (this follows from $\Omega$ being equal for all modes), so the measurement provides no information about the differential momentum. Furthermore, every mechanical mode is driven by the same radiation pressure noise, resulting in no net backaction on the differential momentum; contributions from every odd mode {($p_{1}$, $p_{3}$, ...)} are exactly cancelled by the even modes ($p_{2}$, $p_{4}$, ...).  As a result, the conditional momentum of the differential mode is equal to its unconditional (thermal) momentum,  ${V_{\delta \pre{p} \delta \pre{p}}^{(N-)}}=n_{\text{th}} + 1/2$. The criterion ${V_{\delta \pre{q} \delta \pre{q}}^{(N)} V_{\delta \pre{p} \delta \pre{p}}^{(N-)}<1/4}$ is a sufficient criterion for entanglement~ \cite{bowen_quantum_2015}. Using the thermal momentum and Eqn~\ref{Eqns:Common_position_variance}, this leads to the condition for the two collective modes to become conditionally entangled stated in the main text.

\bibliography{Exp}

\section*{Acknowledgements}
This work was performed in part at the Queensland node
of the Australian National Fabrication Facility, a company
established under the National Collaborative Research
Infrastructure Strategy to provide nano- and microfabrication
facilities for Australia’s researchers. We acknowledge
the facilities, and the scientific and technical assistance,
of the Australian Microscopy \& Microanalysis Research
Facility at the Centre for Microscopy and Microanalysis,
The University of Queensland.
We would like to thank E.H.H. Cheng, G.I. Harris and C.G. Baker for useful discussion. This research was supported by the Australian Research Council Centre of Excellence for Engineered Quantum Systems (EQUS, CE170100009). E.M.B. thanks EQUS for support and funding from an EQUS Deborah Jin Fellowship.

\section*{Author Contributions}
The core concept was conceived by W.P.B., with refinements by C.M. and G.A.B. Structurally-engineered double-disk design and fabrication was performed by G.A.B. with input from C.M. and W.P.B. The optomechanical measurement apparatus was designed by G.A.B. and W.P.B., with C.M. and E.M.B. performing assembly and testing. C.M. collected the measurement data, with important contributions from G.A.B and S.K. Data analysis methods were refined by C.M., G.A.B., J.S.B., S.K., and W.P.B. Analytical models were found by C.M., G.A.B., S.K., J.S.B. and W.P.B. The manuscript was primarily written by C.M., J.S.B., and W.P.B., with contributions from all authors. Figures were chiefly prepared by C.M. Primary supervision for this project was performed by W.P.B., with assistance from J.S.B. and E.M.B.

\section*{Additional information}

The authors declare no competing interests.

The corresponding author is responsible for submitting a \href{http://www.nature.com/srep/policies/index.html#competing}{competing interests statement} on behalf of all authors of the paper. This statement must be included in the submitted article file.

\clearpage

\pagebreak

\begin{center}
	\textbf{\large Supplementary Information: Measurement-based preparation of non-Markovian and multimode mechanical states
}

\end{center}
This Supplementary Information provides details about our filter creation methods, the definition and the role of relative estimates, the existence of structural damping in our optomechanical device  and its effect on the estimation and, finally, the optomechanical parameters of our device.
\setcounter{equation}{0}
\setcounter{figure}{0}
\setcounter{table}{0}
\setcounter{page}{1}
\makeatletter
\renewcommand{\theequation}{S\arabic{equation}}
\renewcommand{\thefigure}{S\arabic{figure}}
\renewcommand{\thepage}{S\arabic{page}}

\section{Causal and anti-causal Wiener filters }

The optimal choice of filter for a linear, continuous measurement is the Wiener filter, which minimises the mean squared error~\cite{wiener_extrapolation_1964}. The causal Wiener filters for estimating  position and  momentum of a single-mode mechanical oscillator with viscous damping have been analytically obtained~\cite{meng_mechanical_2020} and, further, have been directly employed  in an optomechanical experiment~\cite{matsumoto_preparing_2020}. 


The causal Wiener filter for position has been expressed in Eqn \ref{eqns:WienerQ}. The analogous filter function for momentum is
\begin{equation}
	\label{eqns:WienerP}
	\overrightarrow{H}_p(\omega)=\frac{1}{M_{Y}}\left[\frac{S_{p Y}}{M_{Y}^{*}}\right]_{+}.
\end{equation}
%
%
%
Applying the Wiener filters to the measurement record, we have the conditional power spectra
\begin{equation}
    \begin{aligned}
    S_{\delta \pre{q} \delta \pre{q}}=& S_{qq}+S_{\pre{q}\pre{q}}-2\text{Re}\{S_{q \pre{q}}\},
    \\
    S_{\delta \pre{p} \delta \pre{p}}=& S_{pp}+S_{\pre{p}\pre{p}}-2\text{Re}\{S_{p \pre{p}}\},
    \\
    S_{\delta \pre{q} \delta \pre{p}}=& \text{Re}\{S_{YY}\pre{H}_q [\pre{H}_p]^*\}.
    \end{aligned}
\end{equation}
%
%
%
%
Then the position and momentum variances of the conditional state are \cite{brown_introduction_1996}
\begin{equation}
\begin{aligned}
V_{\delta \pre{q} \delta \pre{q}}=&\int_{-\infty }^{+\infty}S_{\delta \pre{q} \delta \pre{q}}\frac{\text{d}\omega }{2\pi}
\\
V_{\delta \pre{p} \delta \pre{p}}=&\int_{-\infty }^{+\infty}S_{\delta \pre{p} \delta \pre{p}}\frac{\text{d}\omega }{2\pi}
\end{aligned}
\end{equation}
Similarly, the covariance $C_{\delta\pre{q} \delta\pre{p}}$ is \cite{brown_introduction_1996}:
\begin{equation}
C_{\delta \pre{q} \delta \pre{p}}=\int_{-\infty}^{+\infty}S_{\delta \pre{q} \delta \pre{p}}\frac{\text{d}\omega}{2\pi}.
\end{equation}
%
The conditional covariance matrix is defined by
\begin{equation}
\overrightarrow{\mathbb{V}}=\left(\begin{array}{ll}
V_{\delta \overrightarrow{q} \delta \overrightarrow{q}} & C_{\delta \overrightarrow{q} \delta \overrightarrow{p}} \\
C_{\delta \overrightarrow{q} \delta \overrightarrow{p}} & V_{\delta \overrightarrow{p} \delta \overrightarrow{p}}
\end{array}\right),
\end{equation}
%
%

The causal filters use the past measurement record to estimate the oscillator's position and momentum. In order to have an estimation based on the future measurement record, we use anti-causal Wiener filters given by
\begin{equation}
\begin{aligned}
	\label{eqns:WienerAnti}
	\overleftarrow{H}_q (\omega)&=\frac{1}{M_{Y}^*}\left[\frac{S_{q Y}}{M_{Y}}\right]_{-},\\
	\overleftarrow{H}_p (\omega)&=\frac{1}{M_{Y}^*}\left[\frac{S_{p Y}}{M_{Y}}\right]_{-}.
\end{aligned}
\end{equation}
$\left[\cdots\right]_{-}$ indicates taking the anti-causal part of the contained function. Following a similar process, we can obtain the retrodicted covariance matrix,
\begin{equation}
\overleftarrow{\mathbb{V}}=\left(\begin{array}{ll}
V_{\delta \overleftarrow{q} \delta \overleftarrow{q}} & C_{\delta \overleftarrow{q} \delta \overleftarrow{p}} \\
C_{\delta \overleftarrow{q} \delta \overleftarrow{p}} & V_{\delta \overleftarrow{p} \delta \overleftarrow{p}}
\end{array}\right).
\end{equation}

Comparing the covariance matrices for the causal and anti-causal filters, we find that  
\begin{equation}
\begin{aligned}
	V_{\delta \overrightarrow{q} \delta \overrightarrow{q}}&=V_{\delta \overleftarrow{q} \delta \overleftarrow{q}},
	\\
	V_{\delta \overrightarrow{p} \delta \overrightarrow{p}}&=V_{\delta \overleftarrow{p} \delta \overleftarrow{p}},
	\\
	C_{\delta \overrightarrow{q} \delta \overrightarrow{p}}&=-C_{\delta \overleftarrow{q} \delta \overleftarrow{p}},
\end{aligned}
\end{equation}
\textit{i.e.}, the causal and anti-causal estimations yield the same error variances but with opposite covariance~\cite{tsang_quantum_2009}.  

\section{The Wiener filters in simulations and experiments 
}
\label{section:simulation_method}

To condition the mechanical state in more complicated circumstances, such as structural damping and multiple modes, we construct the Wiener filters using a numerical approach, as follows, and use them in the simulations and experiments. Firstly, we convert the experimental data from the time domain to the frequency domain, yielding an experimental photocurrent PSD. We then perform a fit to this PSD, assuming that the mechanical contributions are of the form predicted by standard optomechanical models (\textit{i.e.}, $4\eta \mu_j S_{q_{j} q_j}$). Note that all the mechanical modes are assumed to be structurally damped. The fitted parameters are used to calculate the smoothed (idealised) photocurrent PSD and the cross-power spectral densities  as
\begin{equation}
    \begin{aligned}
       \tilde{S}_{YY}&=4 \eta \sum_{j=1}^{N} \mu_j S_{q_j q_j} + S_{NN},
       \\
       S_{qY}&=2\sqrt{\eta}\sum_{j=1}^{N}\sqrt{\mu_j} S_{q_j q_j},
       \\
       S_{pY}&=- i \omega S_{qY},
    \end{aligned}
\end{equation}
where $S_{NN}$ includes  shot noise (modelled as a constant contribution of 1/2), mechanical peaks that are being considered as measurement noise, optical phase noise, and broad-band electronic noise. In this circumstance, the analytical expressions for $M_Y$ and $M_Y^*$ cannot be found. Therefore we numerically find the solutions using the Wiener–Hopf decomposition~\cite{fornberg_complex_2020}. Finally, these can then inserted into Eqns~\ref{eqns:WienerQ}, \ref{eqns:WienerP} and \ref{eqns:WienerAnti}. This produces the frequency domain filters, ready to be applied to the photocurrent to obtain the estimate. The same filters are applied regardless of whether the photocurrent is from the simulation or experiment.



\section{Relative estimates and conditional variances}

The relative estimate variances for position and momentum are defined as
\begin{equation}
\begin{aligned}
\label{Eqns:rela_est_var}
V_{\Delta q \Delta q}&=\langle(\pre{q}-\retro{q})(\pre{q}-\retro{q})\rangle=V_{\pre{q} \pre{q}}+V_{\retro{q}\retro{q}}-2 V_{\pre{q}\retro{q}},
\\
V_{\Delta p \Delta p}&=\langle(\pre{p}-\retro{p})(\pre{p}-\retro{p})\rangle=V_{\pre{p} \pre{p}}+V_{\retro{p}\retro{p}}-2 V_{\pre{p}\retro{p}},
\\
C_{\Delta q \Delta p} & = \langle(\pre{q}-\retro{q})(\pre{p}-\retro{p})\rangle=V_{\pre{q} \pre{p}}+V_{\retro{q}\retro{p}}-2 V_{\pre{q}\retro{p}},
\end{aligned}
\end{equation}
We firstly analyse this for a viscously-damped single-mode mechanical oscillator analytically. The causal ~\cite{meng_mechanical_2020} and anti-causal filter functions are
\begin{equation}
\begin{aligned}
    \pre{H}_{q}(\omega)=&A(1-i B \omega) \chi^{\prime}(\omega),
    \\
    \pre{H}_{p}(\omega)=&-\frac{A B}{\Omega}\left(\Omega^{2}+i \omega \frac{\Omega^{\prime 2}-\Omega^{2}}{\Gamma^{\prime}+\Gamma}\right) \chi^{\prime}(\omega),
    \\
    \retro{H}_{q}(\omega)=&\left[\pre{H}_{q}(\omega)\right]^*,
    \\
    \retro{H}_{p}(\omega)=&-\left[\pre{H}_{p}(\omega)\right]^*,
\end{aligned}
\end{equation}
where $A=8 \sqrt{\eta \Gamma^{3} C} n_{\mathrm{tot}} \Omega^{2} /\left(\Omega^{2}+\Omega^{\prime 2}\right)$ and $B=\left(\Gamma+\Gamma^{\prime}\right) /\left(\Omega^{\prime 2}-\Omega^{2}+\Gamma^{2}+\Gamma \Gamma^{\prime}\right)$ are frequency independent coefficients, and $\Omega^{\prime}=\left(16 \eta \Gamma^{2} C n_{\text {tot }} \Omega^{2}+\Omega^{4}\right)^{1 / 4}$ and $\Gamma^{\prime}=\left(-2 \Omega^{2}+\Gamma^{2}+2 \Omega^{\prime 2}\right)^{1 / 2}$ are the resonance frequency and decay rate in the modified susceptibility $\chi^{\prime}(\omega)=1 /\left(\Omega^{\prime 2}-\omega^{2}-i \Gamma^{\prime} \omega\right)$ of the filter function.

These filter functions allow us to calculate the first two terms in Eqns~\ref{Eqns:rela_est_var} directly, and the third term
\begin{equation}
\begin{aligned}
    V_{\pre{q}\retro{q}} &= \int_{-\infty}^{+\infty} S_{\pre{q}\retro{q}}\frac{\text{d}\omega }{2\pi}=\int_{-\infty}^{+\infty} S_{YY}\left[\pre{H}_{q}\right]^{*} \retro{H}_q \frac{\text{d}\omega }{2\pi},
    \\
    V_{\pre{p}\retro{p}} &= \int_{-\infty}^{+\infty} S_{\pre{p}\retro{p}}\frac{\text{d}\omega }{2\pi}=\int_{-\infty}^{+\infty} S_{YY}\left[\pre{H}_{p}\right]^{*} \retro{H}_p \frac{\text{d}\omega }{2\pi},
    \\
    V_{\pre{q}\retro{p}} &= \int_{-\infty}^{+\infty} S_{\pre{q}\retro{p}}\frac{\text{d}\omega }{2\pi}=\int_{-\infty}^{+\infty} S_{YY}\left[\pre{H}_{q}\right]^{*} \retro{H}_p \frac{\text{d}\omega }{2\pi},
\end{aligned}
\end{equation}
Inserting these terms in Eqns~\ref{Eqns:rela_est_var} results in
\begin{equation}
\begin{aligned}
V_{\Delta q \Delta q}=&\frac{A^{2} B^{2}}{\Gamma+\Gamma '}\left(1+\frac{2 \Gamma \Gamma '}{\Omega^{2}+\Omega '^{2}}\right),
\\
V_{\Delta p \Delta p}=&\frac{ A^{2} B^{2}}{(\Gamma+\Gamma ')^{2} \Omega^{2}}\left(\frac{\Gamma \Omega^{2}}{2}+\frac{\Gamma ' \Omega '^{2}}{2}+\frac{\Gamma \Gamma '(\Gamma-\Gamma ')\left(\Omega '^{2}-\Omega^{2}\right)}{2\left(\Omega^{2}+\Omega '^{2}\right)}\right),
\\
C_{\Delta q \Delta p} =& 0.
\end{aligned}
\end{equation}
We relate these quantities to the conditional position and momentum variance by introducing the conversion factors $F_q$ and $F_p$:
\begin{equation}
\begin{aligned}
V_{\Delta q \Delta q}&=
2(1-F_q)V_{\delta \pre{q} \delta \pre{q}},
\\
V_{\Delta p \Delta p}&=2(1-F_p)V_{\delta \pre{p} \delta \pre{p}}.
\end{aligned}
\end{equation}
and find that $F_q$ and $F_p$ vanish in the high measurement rate regime~\cite{rossi_observing_2019} for a viscously damped oscillator.



For a more general case, such as our multimode structurally damped oscillator, the conversion factors can be obtained using the simulation approach introduced in section~\ref{section:simulation_method}. These conversion factors allow us to infer the conditional variances from the relative estimate variances in the experiment. In Table~\ref{Table:F}, we have listed the values for our measurements in the case of estimating using a different number of modes. For nine-mode conditioning, the thermal squeezing ratio increases to 2.7 (from 2.3 for five-mode conditioning).

%


\begin{table}[H]
\centering
\begin{tabular}{|c|c|c|c|c|c|c|}
\hline
 $N$ & $V_{\Delta q \Delta q}$/2 & $V_{\Delta p \Delta p}$/2 & $F_q$   & $F_p$    & $V_{\delta \pre{q} \delta \pre{q}}$ & $V_{\delta \pre{p} \delta \pre{p}}$ \\ \hline
1                                               & 1.8$\times 10^5$    & 3.0$\times 10^5$    & 0.26 & -0.11 & 2.5$\times 10^5$    & 2.7$\times 10^5$    \\ \hline
2                                               & 1.0$\times 10^5$    & 2.4$\times 10^5$    & 0.39 & -0.18 & 1.6$\times 10^5$    & 2.0$\times 10^5$    \\ \hline
3                                               & 1.1$\times 10^5$    & 2.5$\times 10^5$    & 0.38 & -0.17 & 1.8$\times 10^5$    & 2.1$\times 10^5$    \\ \hline
4                                               & 1.1$\times 10^5$    & 2.7$\times 10^5$    & 0.32 & -0.09 & 1.6$\times 10^5$    & 2.5$\times 10^5$    \\ \hline
5                                               & 9.1$\times 10^4$    & 3.0$\times 10^5$    & 0.21 & -0.13 & 1.1$\times 10^5$    & 2.6$\times 10^5$    \\ \hline
6                                               & 9.0$\times 10^4$    & 3.1$\times 10^5$    & 0.21 & -0.13 & 1.1$\times 10^5$    & 2.7$\times 10^5$    \\ \hline
7                                               & 8.4$\times 10^4$    & 3.0$\times 10^5$    & 0.21 & -0.13 & 1.1$\times 10^5$    & 2.7$\times 10^5$    \\ \hline
8                                               & 8.7$\times 10^4$    & 3.1$\times 10^5$    & 0.22 & -0.12 & 1.1$\times 10^5$    & 2.8$\times 10^5$    \\ \hline
9                                               & 6.6$\times 10^4$    & 2.7$\times 10^5$    & 0.24 & -0.13 & 8.7$\times 10^4$    & 2.4$\times 10^5$    \\ \hline
\end{tabular}%
\caption{\label{Table:F}Table of the relative estimate variances, conversion factors and conditional variances in the experiment by estimating a different number of modes $N$.}
\end{table}

\section{Thermal noise of structural damping}
\begin{figure}[H]
	\centering
		\includegraphics[width=0.7\textwidth]{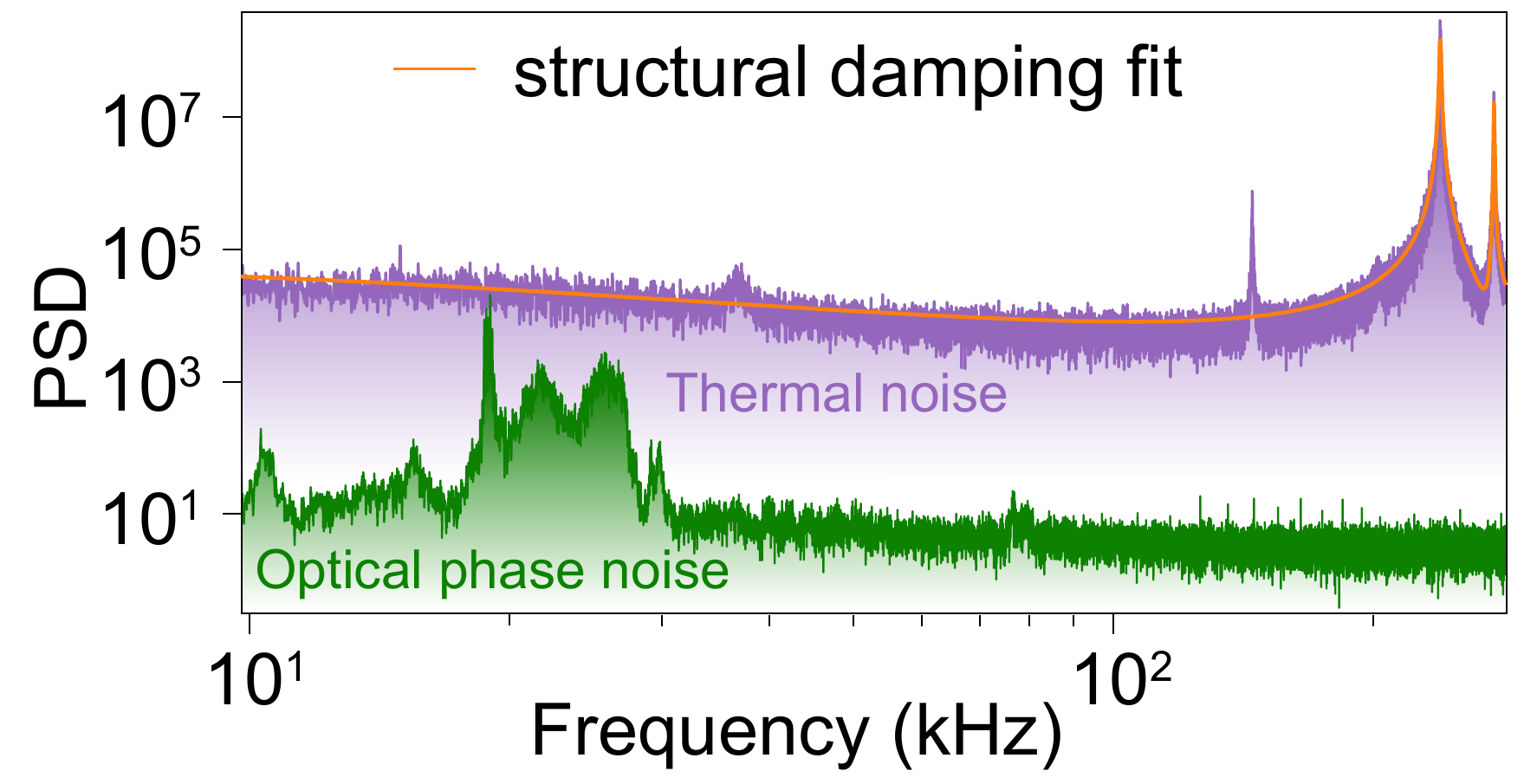}
		\caption{\label{Fig:phase noise} Comparison of the power spectral density of the thermal noise (purple) and an upper bound measurement of the optical phase noise (green). The residual phase noise is attributed to acoustic vibration of the optical fibres. }
\end{figure}

Structural damping – damping in which the loss is approximately independent of frequency – is encountered in many engineered structures~\cite{kimball_internal_1927,saulson_thermal_1990,de_silva_vibration_2007, neben_structural_2012,fedorov_evidence_2018, cripe_measurement_2019,komori_attonewton-meter_2020}. In general, the displacement spectrum of a mechanical oscillator is given by
\begin{equation}
\label{eqns:structural_damping}
S_{qq}=\frac{2 n_{\mathrm{th}} \Omega^{2} \phi\left(\omega\right)}{\omega} \frac{\Omega^{2}}{\left(\omega^{2}-\Omega^{2}\right)^{2}+\Omega^{4} \phi^{2}\left(\omega\right)},
\end{equation}
where substituting ${\phi(\omega)={\omega}Q^{-1}/{\sqrt{\omega^{2}+\omega_{c}^{2}}}}$ leads to structural damping and  $\phi(\omega) =\omega Q^{-1}/\Omega $ leads to viscous damping. The thermal noise PSD of the structurally damped oscillator with $1/f$ feature is shown in Fig.~\ref{Fig:phase noise}. In order to confidently attribute this $1/f$ noise to structural damping, we take additional measurements to rule out other sources of noise, including electronic noise and laser phase noise. Our measurement set-up is already designed to minimise laser phase noise by matching the path length of the two arms of the interferometer (shown in Fig~\ref{Fig:Diagram}\textbf{c}) such that correlated laser phase noise cancels out. The path lengths are matched to within 0.5~m while taking the measurements for conditioning. To make an upper-bound measurement of the laser phase noise contribution we uncouple the optical fibre from the optomechanical device, which acts to increase the path length mismatch of the interferometer by the effective cavity length (approximately 0.5~m), and we make a measurement with everything else remaining identical. The results of this measurement are shown in Fig~\ref{Fig:phase noise}, confirming that the optical phase noise is significantly below the level of the $1/f$ noise attributed to structural damping, in addition to having a different spectral shape.

\end{document}